# Development of an Interatomic Potential for the Simulation of Defects, Plasticity and Phase Transformations in Titanium

M.I. Mendelev[1*], T.L. Underwood[2] and G.J. Ackland[3]

[1]Division of Materials Sciences and Engineering, Ames Laboratory, Ames, IA, 50011, USA
[2] Department of Physics, University of Bath, Bath, BA2 7AY, UK
[3]Centre for Science at Extreme Conditions, School of Physics, University of Edinburgh, Edinburgh EH9 3JZ, Scotland, UK

**Abstract**
New interatomic potentials describing defects, plasticity and high temperature phase transitions for Ti are presented. Fitting the martensitic hcp-bcc phase transformation temperature requires an efficient and accurate method to determine it. We apply a molecular dynamics (MD) method based on determination of the melting temperature of competing solid phases, and Gibbs-Helmholtz integration, and a lattice-switch Monte Carlo method (LSMC): these agree on the hcp-bcc transformation temperatures to within 2 K. We were able to develop embedded atom potentials which give a good fit to either low or high temperature data, but not both. The first developed potential (Ti1) reproduces the hcp-bcc transformation and melting temperatures and is suitable for the simulation of phase transitions and bcc Ti. Two other potentials (Ti2 and Ti3) correctly describe defect properties, and can be used to simulate plasticity or radiation damage in hcp Ti. The fact that a single EAM potential cannot describe both low and high temperature phases may be attributed to neglect of electronic degrees of freedom, notably bcc has a much higher electronic entropy. A temperature-dependent potential obtained from the combination of potentials Ti1 and Ti2 may be used to simulate Ti properties at any temperature.

[*] Corresponding author; email address: mendelev@ameslab.gov

# 1. Introduction

Phase transformations play a key role in many material manufacturing techniques and the ability to control them is a path to achieving desired material properties. Understanding the atomistic mechanism of phase transformation is an obvious prerequisite to control them. Atomistic computer simulation can be especially valuable to gain insight into details of phase transformation. The most reliable simulations involve *ab initio* calculations to find all energies at $T=0$. However, since most phase transformations take place at high temperatures the values obtained at $T=0$ may not be very relevant. Moreover many material properties of interest, from plasticity to radiation damage, are governed by collective atomic-level processes, but occur over length-scales beyond what is tractable using *ab initio* methods. An alternative approach is to fit a semi-empirical potential to the *ab initio* data at T=0 and employ it in molecular dynamics (MD) simulation of the process under investigation [1]. In this paper, using Ti as an example, we discuss how to develop and verify such semi-empirical potentials.

Titanium is among the most important metals for use as a structural material, typically with small amounts of alloying elements. It would therefore seem to be a natural candidate for molecular dynamics simulations to determine the microstructures, machinability and defect properties. However, there are relatively few such studies, primarily because the critical properties of titanium are poorly described by existing interatomic potentials. The two main problematic features are the existence of two solid phases, hcp (α) and bcc (β) and the very high basal stacking fault energy which leads to deformation via prism and twinning slip. To our knowledge, these effects are not correctly described by current potentials. Semi-empirical potentials tend to predict rather low basal stacking fault energies in Ti [2-6] .

A further problem is that a very large number of *ab initio* calculations all predict that the lowest energy structure is the complex ω phase (e.g., see [5, 7, 8]). Titanium has four valence electrons, so the binding has both s- and d-like character. Furthermore, the bcc structure is unstable at low temperature (it has a negative elastic constant $C'=C_{11}-C_{12}$), and the contribution from the electronic entropy is significant. Thus an interatomic potential needs to describe three effects: i) free-electron like s-bonding, which suggests long ranged pairwise interactions with Friedel oscillations; ii) tightly bound d-bonding, which suggests a second moment tight binding Finnis-Sinclair [9] or an embedded atom method (EAM) [10]-type model and iii) temperature dependence arising from the electronic entropy. An additional challenge arising in developing a semi-empirical potential for Ti is its rather high values of the stacking fault energies.

The main thermodynamic parameters describing a first order phase transformation are the transformation temperature, the latent heat and the change in the atomic volume. It is relatively straightforward to incorporate the latent heat and the change in the atomic volume in the potential development procedure (see Section 2). A method to fit the transformation temperature was proposed in [11]. In this method, the parameters of the phase transformation are calculated using a trial semi-empirical potential and then the Gibbs-Duhem equation is used to correct the semi-empirical potential parameters to achieve a better agreement with the target (experimental) phase transformation parameters. The authors of [11] demonstrated that this method can be used to correct the melting temperature of an Al EAM potential without considerable change in other material properties. The method proposed in [11] can be easily incorporated in the potential development procedure to construct semi-empirical potentials to simulate the solid-liquid interface properties (e.g., see [12-14]). As follows from the derivation of this method in [11], it should be equally possible to use the method to adjust a semi-empirical potential to the temperature of a solid-solid phase transformation. However, an obvious difficulty is that while it is rather easy to determine the melting temperature from MD simulation (e.g., see [15, 16]), the determination of the solid-solid transformation temperature is challenging because of relatively low interface mobility in such transformations. If both solid phases can co-exist with a liquid during MD simulation the solid-solid transformation temperature can be determined from their the melting temperatures (e.g., see [12, 17]). However, to our best knowledge, the accuracy of this procedure was never determined by a comparison with results from an independent technique.

A number of practicable methods do exist which allow solid-solid transformation temperatures to be determined directly, the most prominent of these being the quasiharmonic method [18, 19] and thermodynamic integration [20, 21]. Both of these methods involve explicitly calculating the free energies of the two solid phases under consideration as a function of T, and determining the T at which the free energies are equal – which is the transformation temperature. In the quasiharmonic method the harmonic approximation is assumed to apply separately to each volume of the system; the dynamical matrix and phonon density of states are volume-dependent. Here the free energy for a given phase is obtained by first calculating its phonon density of states $g(\omega,V)$ as a function of volume, $V$, and then exploiting well-known expressions relating $g(\omega,V)$ and $T$ to the free energy. However, while the quasiharmonic method has proved successful, its underlying assumptions can break down at high temperatures where high-order anharmonic effects, not accounted for in the method, become important. In the case of dynamically stabilized materials such as bcc Ti, the imaginary phonon modes at T=0 lead to divergent free energy. Thermodynamic integration does not suffer from this deficiency, being in principle exact for all temperatures. In this method a pathway is constructed linking the system under consideration to a 'reference system' whose free energy is known. The free energy of the system under consideration is then calculated by integrating the free energy along the pathway. The reference system need not be physical in the sense that it can have a different Hamiltonian to the 'real' system, in which case the pathway involves changing the Hamiltonian of the system. In fact this is the most powerful incarnation of thermodynamic integration, and has been widely applied.

The downside of thermodynamic integration is that, compared to the quasiharmonic method, it is computationally expensive. This is problematic because the efficiency of the method one uses to determine the transformation temperature has a bearing on how precisely it can be determined. With this in mind a number of methods which can treat highly-anharmonic solids have been developed. These include methods which extend the quasiharmonic approximation to account for additional anharmonic effects, such as self-consistent *ab initio* lattice dynamics [22], the inverse Z-method [23] and lattice-switch Monte Carlo (LSMC) simulation [24-26]. LSMC, like thermodynamic integration, is in principle exact, and its accuracy has been well established: LSMC has been tested against the harmonic approximation for the Lennard-Jones solid [26], and against thermodynamic integration for various soft-matter systems [27, 28]. Furthermore LSMC is ostensibly the most computationally efficient method [27, 29] for calculating solid-solid free energy differences (though this claim has been disputed [28]). For this reason using LSMC to calculate solid-solid transformation temperatures with the aim of developing semi-empirical potentials, or cross-checking other methods, is an interesting prospect.

The rest of the paper is organized as follows. First, we describe the potential development procedure and present several new Ti EAM potentials. Next, we will discuss two methods to determine a solid-solid transformation temperature and show that they lead to essentially the same results. One is MD-based, deducing the transformation temperature indirectly from the melting temperatures of both solid phases as mentioned above. The other method is LSMC-based. Finally, we discuss the advantages and deficiencies of developed Ti EAM potentials.

## 2. Potential development procedure

The total energy in a single component system described by an EAM potential takes the following form:

$$U = \sum_{i=1}^{N-1} \sum_{j=i+1}^{N} \varphi(r_{ij}) + \sum_{i=1}^{N} F(\rho_i) \,, \tag{1}$$

where the subscripts i and j label each of the $N$ atoms in the system, $r_{i,j}$ is the separation between atoms i and j, $\varphi(r)$ describe the pair interaction, $F(\rho_i)$ is the embedding energy contribution which depends on the "electron density" of the atom $i$:

$$\rho_i = \sum_j \psi(r_{ij}) \,, \tag{2}$$

where $\psi(r)$ is the "density function". Thus, an EAM potential contains three functions ($\varphi(r)$, $F(\rho)$ and $\psi(r)$) and developing an EAM potential involves optimization of these functions.

Interatomic potentials can be used for many applications, and in a given case, some properties may be more crucial than others. In the case of titanium, one may be interested in the high temperature behavior for microstructure formation, or low temperature behavior for mechanical applications. Thus we fit different potentials with these cases in mind.

Four groups of target properties were used in the potential development procedure. The first group includes basic T=0 perfect crystal properties listed in Table I. Fitting to these properties ensures that hcp is the most stable phase at T=0 and sets the correct energy scale. It should be noted that according to the *ab initio* calculations the Ti bcc phase is mechanically unstable at T=0 ($C_{11} < C_{12}$) [5, 7, 30, 31]. Therefore, reproducing the target values at T=0 for this phase does not ensure the correct description of this phase at high temperatures (including its mechanical stability).

The properties in the first group are routinely included in any potential development procedure. The only complexity in the case of hcp metals is associated with the fact that the elastic constants cannot be determined via the virial expressions and molecular static relaxation is required for their determination (which was not taken into account in [12]). The potential development procedure was adjusted to account for this fact.

The second group of properties (Table II) includes point defect formation energies. Fitting to these data is supposed to make the semi-empirical potential suitable for simulation of self-diffusion and other processes which atomic jumps are involved. A special feature of hcp metals is the existence of multiple self-interstitial locations [32]. It is very difficult to fit a semi-empirical potential to correctly reproduce all of these energies. Therefore, we mostly focused on correct reproduction of the most stable self-interstitial configurations (O, BO and BS – see Fig. 1 in [32]). Since the bcc phase is unstable at T=0 we did not include its point defect energies in the potential development procedure.

The third group of properties (Table III) includes planar defect formation energies. In the present work, a special attention was paid to correct reproduction of the stacking fault energies in hcp which is important for simulation of the plastic deformation. The free surface was not the main focus and we tried to provide a reasonable value for this quantity rather than to exactly reproduce it.

Finally, the fourth group of properties (Table IV) includes the phase transformation data. In the present study we considered three phase transformations: melting of the hcp and bcc phases and hcp-bcc transformation. The only experimental information we found for these transformation was the bcc melting temperature, the hcp-bcc transformation temperature and the latent heats for these transformations. There is indirect evidence that bcc Ti phase is more dense than the hcp phase [33] however, the experimental data were obtained at room temperature. Therefore, no data on the change in the atomic volume upon phase transformations were included in the potential development procedure.

To fit the phase transformation temperature we used the following equation for the parameters of a semi-empirical potential, $a_k$, derived in [11]:

$$T^{cur}_{\alpha \to \beta} + \sum_k T^{exp}_{\alpha \to \beta} \frac{\lambda^{\beta}_k - \lambda^{\alpha}_k}{U^{\beta} - U^{\alpha}} \left( a_k - a^{cur}_k \right) = T^{exp}_{\alpha \to \beta}, \qquad (3)$$

where $T^{exp}_{\alpha \to \beta}$ is the experimental phase transformation temperature, $T^{cur}_{\alpha \to \beta}$ is the phase transformation temperature obtained with the current semi-empirical potential with parameters $a^{cur}_k$, $U^{\alpha}$ is the potential energy of the phase $\alpha$, and

$$\lambda^{\alpha}_k = \left\langle \left( \frac{\partial U^{\alpha}}{\partial a_k} \right)_{a_{l \neq k}} \right\rangle_{N,p,T}. \qquad (4)$$

Obviously, this method implicitly assumes also fitting the latent heat ($U^{\beta}-U^{\alpha}$) and the atomic densities of the co-existing phases (or at least both phases should be treated at the same pressure). Fitting the energy and atomic density of a crystal phase is rather straightforward because the structure is well defined by the crystal symmetry. Fitting the energy and atomic density of a liquid phase is more complicated because refitting a semi-empirical potential can lead to a different liquid structure. To avoid this problem we added constraints on the liquid structure: it could not considerably change between two consequent iterations of the potential development procedure. To do this we used the method to fit the liquid structure proposed in [34]. Finally, application of Eq. (3) requires knowledge of the phase transformation temperature for the current semi-empirical potential. The methods to determine the phase transformation temperatures will be discussed in Section 3.

Most of the target properties listed in Tables I-IV were used in the fitting procedure but with different weight. The developed potential functions are shown in Fig. 1. All potentials have the same density function but different pair potential and embedding energy functions. The potentials can be found in [35] and [36].

The main objective in developing potential Ti1 was to reproduce the phase transformation data. The potential very well reproduces both the melting temperature and hcp-bcc transformation temperatures however it underestimates the latent heat. It does predict that the bcc phase is denser than the hcp phase at the hcp-bcc transformation temperature. This was achieved by making considerable compromise in reproduction of other target properties. The most obvious deficiency is the hcp vacancy formation energy which is higher than the interstitial formation energy (this issue will be discussed in Section 5). The order of magnitude of the lowest energy self-interstitials is reasonably reproduced but the self-interstitials O and BO have about the same energy while *ab initio* calculations predict that self-interstitial O has smaller energy.

To determine the finite and high temperature properties, we first performed a careful determination of the hcp, bcc and liquid bulk properties corresponding to each of the EAM potentials developed within the present study. In order to determine the equilibrium atomic density we used simulation cells containing 2000 (bcc, liquid) atoms. At a particular temperature, an *NVT* (constant number of atoms, volume and temperature) MD simulation was performed at several densities for 20,000 MD steps (~40.7 ps) and then the pressure was averaged over the next 40 ps. Using the obtained dependences of the pressure, *p*, on atomic volume, *p(V)*, the equilibrium density was found from the condition *p(V)=0*. The simulation cell was then equilibrated at the equilibrium density for 40 ps and the total energy was averaged over the next 40 ps. In the case of the hcp phase the simulation cell contained 1848 atoms and NVT simulation were run at 4 combinations of the lattice parameters *a* and *c*. The equilibrium lattice parameters were obtained assuming the linear elasticity regime. The average values of stresses obtained during the final NVT simulation at equilibrium lattice parameters did not exceed 0.04 GPa which corresponds the error in the determination of the lattice parameter not larger than 0.001 Å.

The hcp lattice parameters as functions of temperature are shown in Fig. 2. The lattice parameter *a* monotonically increases with increasing temperature. By contrast, the hcp lattice parameter *c* at very low temperature slightly decreases with increasing temperature reaching a minimum at 90 K and then it increases with increasing temperature. It should be noted that overall the atomic volume monotonically increases with increasing temperature. Potential Ti1 provides a very good agreement with the experimental data [37] on the thermal expansion for the hcp lattice parameter, *a*, but considerably underestimates the thermal expansion of *c*. Therefore, this potential leads to a monotonic decrease in the c/a ratio with increasing temperature while the experiment shows that this ratio is almost temperature independent [37].

The main objective of developing potential Ti2 was a better description of the point defect formation energies in the hcp phase. It was found that using the present potential form this goal can be achieved only by excluding the bcc melting temperature from the list of target properties. As a result, potential Ti2 dramatically underestimates the bcc melting temperature but still leads to the correct hcp-bcc transformation temperature. Table II demonstrates that the hcp point defect formation energies provided by this potential are in very good agreement with the *ab initio* data.

Similarly to potential Ti1, the hcp lattice parameter *a* calculated with potential Ti2 monotonically increases with increasing temperature while the hcp lattice parameter *c* at low temperature decreases with increasing temperature reaching a minimum at 180 K and then it increases with increasing temperature. To fix this deficiency we developed potential Ti3. This potential provides monotonic increase in both hcp lattice parameters with increasing temperature and almost temperature independent c/a ratio in agreement with experiment. However, potential Ti3 predicts a positive change in the atomic volume upon the hcp-bcc transformation and lead to slightly larger disagreement with the ab initio data on the point defect formation energies.

## 3. Stability of the bcc phase

As was mentioned above, the ratio of the elastic constants, $C_{11}/C_{12}$ for the Ti bcc phase is less than 1 at T=0 [5, 7, 30, 31]. Therefore, even excellent reproduction of the *ab initio* data on the bcc lattice parameter and relative (to hcp) formation energy at T=0 does not really provide any reliability of a semi-empirical potential at high temperature. Therefore, we did not include these data in the potential development procedure. Instead, we were focused on the satisfactory reproduction of the bcc melting temperature and latent heat in the case of potential Ti1 and the hcp-bcc transformation temperature and latent heat in all cases.

Figure 3 shows the bcc lattice parameter as function of temperature. In all cases the lattice parameter monotonically increases with increasing temperature. Potential Ti1 provides an excellent agreement with the experimental data from [37] and potentials Ti2 and Ti2 overestimate the bcc lattice parameter (though they provide reasonable thermal expansions). All potentials lead to the bcc energy higher than the hcp energy but the value of the difference is very different. The data for potential Ti3 are in the best agreement with both T=0 *ab initio* calculations and experimental data on the latent heat of the hcp-bcc phase transformation.

To find the range of the mechanical stability of the bcc phase we determined the $C_{11}/C_{12}$ ratio from MD simulation using the procedure described in [38]. The obtained $C_{11}/C_{12}$ ratio as function of temperature is shown in Fig. 4. For all developed potentials this ratio rather abruptly drops near by a critical temperature. The bcc phase is a metastable phase only above this critical temperature. Its values are rather different for the developed potentials: ~185 K, 500 K and 680 K for Ti1, Ti2 and Ti3, respectively.

## 4. Determination of the phase transformation temperatures

The fact that the bcc phase is mechanically unstable at low temperature makes it difficult to apply the harmonic approximation to determine the hcp-bcc transformation temperature [39, 40]. Therefore, in the present study we applied two other approaches to determine this temperature. Of course, each of them would be sufficient for the purpose of developing a potential but to our best knowledge the reliability of these approaches were never carefully studied by comparison of their results for the same EAM potential.

*4.1. Determination of the phase transformation temperatures from MD integration*

The method to determine the hcp-bcc transformation temperature was proposed in [12, 17]. This method is based on careful determination of the melting temperatures of both crystal phases, $T_m^\alpha$, and using the Gibbs-Helmholtz relation:

$$\int_{T_{hcp \to bcc}}^{T_m^{bcc}} \frac{\Delta U_{hcp \to bcc}}{T^2} dT + \int_{T_m^{bcc}}^{T_m^{hcp}} \frac{\Delta U_m^{hcp}}{T^2} dT = 0. \qquad (5)$$

where the latent heats of the hcp-bcc transformation and hcp melting ($\Delta E^{hcp \to bcc}(T)$ and $\Delta E_m^{hcp}(T)$) are directly determined from the molecular dynamics. The application of this method requires an accurate determination of the melting temperature. In the present study we used the same technique as described in [14] which allows to obtain the melting temperature with inaccuracy not exceeding 0.05% of its value.

Once the transformation temperature has been determined, it is straightforward to obtain the difference in the free energy and entropy between the hcp and bcc phases. The results are shown in Fig. 3. While the trend is the same for all developed potentials the obtained values are strongly dependent on the employed potential. Near the transformation temperature the most realistic values are probably provided by potential Ti3 which leads to the best agreement with the experimental data on the latent heat.

*4.2. Lattice-switch Monte Carlo calculations*

We have also used the lattice-switch Monte Carlo (LSMC) method [24-26] to closely examine the hcp–bcc transformation for the potentials developed in this work. LSMC is an exact method which provides properties pertaining to two solid phases under given conditions, in particular their free energy difference. By 'exact' here we mean that LSMC relies upon no approximations beyond those present in the representation of the system to which it is applied: LSMC calculations will converge upon the exact results for a given potential and given system size. The key feature of LSMC is a lattice-switch Monte Carlo move which transforms a microstate associated with one phase into a microstate associated with the 'other' phase. Lattice-switch moves, if successful, take the system directly from one phase to the other, eliminating the need for the system to traverse any free energy barrier which separates the two phases. Such a barrier prevents both phases from being explored in a tractable amount of simulation time via conventional methods. By using lattice-switch moves to transformation between phases in conjunction with 'conventional' Monte Carlo moves (i.e., atom translation moves and, in the case of the isothermal–isobaric ensemble, volume moves; see, e.g., [20]) to sample microstates within each phase, LSMC can sample both phases in a single simulation of reasonable length. This allows the free energy difference between the phases to be determined directly, and hence ostensibly more efficiently, and hence to a higher precision, than alternative 'exact' methods [27, 29].

While LSMC provides a means of calculating the free energy difference between the two phases at a given temperature and pressure, there is the question of how one uses LSMC to determine the transformation temperature, i.e., the temperature at which the free energy difference is zero. In the present study we used the following iterative procedure which in [26] was shown to be more efficient than alternative methods. We first performed an LSMC calculation at an initial guess $T^{(1)}$ for the transformation temperature. This calculations yielded the enthalpies and volumes of the hcp and bcc phases ($H_{hcp}$, $H_{bcc}$, $V_{hcp}$ and $V_{bcc}$, respectively), the Gibbs free energy difference between the phases $\Delta G = G_{bcc} - G_{hcp}$, as well as associated uncertainties in these quantities at $T^{(1)}$. We then used the following equation to generate a more accurate estimate for the transformation temperature, $T^{(2)}$:

$$T^{(n+1)} = T^{(n)} \left[ 1 - \left( 1 - \frac{\Delta H^{(n)}}{\Delta G^{(n)}} \right)^{-1} \right] \quad (6)$$

We then repeated all of the above, performing LSMC calculations at $T^{(2)}$, $T^{(3)}$, etc., until a sufficiently accurate estimate of the transformation temperature was obtained. We considered this to be the case if $\Delta G^{(n)} = 0$ to within the uncertainty in $\Delta G^{(n)}$, at which point the transformation quantities $\Delta H_{hcp \to bcc}$ and $\Delta V_{hcp \to bcc} / \Delta V_{hcp}$ were obtained from the results of the final iteration $n$.

Technical details and full results of our LSMC calculations are provided in [35]. Here we just note that the simulation cell used in the LSMC contained only 384 atoms. One could argue that a bcc phase could have relatively soft low frequency shear modes with nonlinear dispersion which could lead to a relatively large size effect and, therefore, that a small simulation cell could not be sufficient for such calculations. To test this concern we also performed an LSMC calculation for a simulation cell containing 1296 atoms for potential Ti2 (see [35] for details). We found that our 1296-atom results were statistically indistinguishable from our 384-atom results, indicating that our 384-atom results are converged with respect to system size. This echoes our previous findings in [41], where we examined the hcp-bcc transformation for Zr potential #2 from [12] via LSMC, and where we also did not find a statistically meaningful difference between results based on simulation cells containing 384 or 1296 atoms.

As our initial guess $T^{(1)}$ we used the transformation temperature obtained from our MD simulations described in the previous section. With this we found that the procedure converged quickly, at $n = 2$ for potential Ti1, and just $n = 1$ for potentials Ti2 and Ti3. To crosscheck the validity of the iterative procedure we also performed a set of calculations for potential Ti2 starting with an initial guess $T^{(1)}$ far from the true transition temperature for this potential, namely $T^{(1)}$=900K. The results obtained in this case did not differ significantly from those obtained using the MD temperature as the initial guess, though more iterations were required to convergence upon the transition temperature: $n=4$ instead of $n=1$.

Our LSMC results are presented in Table IV, where it can be seen that the LSMC and MD simulation results are in excellent agreement. This is striking because the two methods are so different, i.e., small number of atoms (384) and Monte Carlo simulation in the first approach vs. larger number of atoms (~2000) and molecular dynamics simulation in the second approach, and is a testament to the accuracy of both methods. In practice the LSMC requires less computational power. In addition, the method described in Section 4.1 can be applied only in the case when the low temperature phase can co-exist with the liquid phase. Taking into account that the difference between the solid-solid transformation and melting temperatures can be very large (like in the case of real Ti) this condition may not be satisfied. For example, a high temperature solid phase can form at the low-temperature-solid-phase – liquid interface which will make impossible to determine the low temperature solid phase melting temperature.

## 5. Simulation of point defect properties

Point defect properties define the diffusion mechanism. There is a lot of controversy about the self-diffusion mechanism in Ti in the literature. Even the range of activation energy obtained from experiment for hcp Ti is rather large (see Table V) which does not allow the use of experimental results to make any closing conclusion. In theoretical considerations, it is widely accepted that the self-diffusion proceeds through the vacancy mechanism [42-45] although the authors of [42] admitted that the self-interstitial mechanism could "make a significant contribution to self-diffusion". Among the potentials we developed in the present study, potential Ti1 seems to be not suitable for the simulation of the self-diffusion in hcp since it leads to the hcp vacancy formation energy larger than the self-interstitial formation (see Table I). Potentials Ti2 and Ti3 lead to the hcp vacancy formation energy which is smaller than the self-interstitial formation energy. Based on this fact one can assume that these potentials lead to a different diffusion mechanism in hcp Ti. However, the data presented in Table I were obtained at T=0 and may differ from the values at finite temperatures where the self-diffusion actually takes place and the self-diffusion mechanism is defined by the activation energy which is the sum of the point defect formation and migration energies. To determine the point defect properties we used the same techniques as in [38]: one point defect (vacancy or interstitial) was introduced in the simulation cell containing 1848 (hcp) or 2000 (bcc) atoms. The simulation cell was equilibrated during 40 ps using NVT MD simulation and then the energy was averaged over the next 400 ps. These energies were used to determine the point defect formation energies as function of temperature (Fig. 5). Next the effective atomic diffusivities were determined during 10 ns using the standard relation between the diffusivity and the atomic mean square displacement:

$$D^{eff} = \frac{<\Delta r^2>}{6t}. \tag{7}$$

The obtained values fall rather well on straight line in the Arrhenius coordinates (Fig. 6) such that the point defect migration energies were obtained via slopes to these lines.

In the case of hcp Ti the *ab initio* calculations predict the vacancy migration energy ranges from 0.4 to 0.52 eV/atom depending on the employed approximations and the direction of the self-diffusion (see below). Examination of Table V shows that potential Ti1 considerably overestimates the vacancy migration energy while potentials Ti2 and Ti3 provides values in excellent agreement with the *ab initio* calculations.

In the case of hcp Ti, the experiment shows that the diffusivity in the direction parallel to the c-axis is two times smaller than the diffusivity in the perpendicular direction [46]. The *ab initio* calculations made in [45] for the vacancy mechanism of diffusion predicts that the $D_{//}/D_{\perp}$ ratio should vary from 0.33 to 0.5 depending on temperature. The *ab initio* calculations made in [44] also lead to the prediction that the diffusion in the direction parallel to the c-axis is slower but the $D_{//}/D_{\perp}$ ratio was found to be 0.89. Potential Ti1 leads to the prediction that the diffusion by the vacancy mechanism along the c-axis is faster while predictions obtained using potentials Ti2 and Ti3 are in agreement with the *ab initio* calculations (Fig. 7). However, all developed potentials lead to the prediction that the self-diffusion in the interstitial mechanism is slower along the c-axis. Therefore, the $D_{//}/D_{\perp}$ ratio cannot really serve as an indicator of what mechanism actually governs the self-diffusion. Since the anisotropy in diffusivity is much smaller than the difference between the diffusivity from *ab initio* calculations and experiment (which is about an order of magnitude [44]) in our further analysis we will ignore the anisotropy.

Figure 5 shows that the vacancy formation energy increases with increasing temperature. The interstitial formation energy for potential Ti1 slightly increases with increasing temperature, reaches a maximum and then decreases. In the case of potentials Ti2 and Ti3 (which are supposed to better describe the hcp point defect properties) the interstitial formation energy monotonically decreases with increasing temperature such that it becomes even smaller than the vacancy formation energy around the temperature of the hcp-bcc transformation. Taking into account that the interstitial migration energy is much smaller than the vacancy migration energy in hcp Ti (see Table VI) all developed potentials lead to the prediction that the self-diffusion in hcp Ti proceeds through the interstitial mechanism (see the activation energies provided in Table V). Moreover, it should be noted that the self-interstitial formation entropy in the hcp phase should be considerably higher than the vacancy formation entropy because there are several interstitial configurations with about the same energy and only 1 vacancy configuration (see also the estimation made in [38] for the hcp Zr). This should additionally increase the equilibrium self-interstitial concentration and make the interstitial mechanism of the self-diffusion more favorable.

In the case of bcc Ti, only the results obtained with potential Ti1 can be compared with experiment since potentials Ti2 and Ti3 dramatically underestimate the melting temperature. Figure 5 shows that the self-interstitial formation energy is much smaller than the vacancy formation energy such that the self-diffusion should proceeds via the interstitial mechanism. It should be noted that the experimental observations show that in the case of self-diffusion in bcc Ti, the diffusivity does not obey the Arrhenius law: the activation energy increases with increasing temperature [47]. Yet the value of the activation energy for the self-diffusion obtained with potential Ti1 in the interstitial mechanism is in reasonable agreement with the experimental data.

## 6. Discussion

We have developed interatomic potentials for Ti within the EAM formalism which reproduce a wide range of properties more accurately than previous potentials, whether pair-potential, Finnis-Sinclair, EAM, or MEAM. All potentials correctly reproduce the hcp stacking fault energies. These are crucial for any simulation of plasticity, because they determine the number of active slip systems. Five slip systems are needed for plasticity, and titanium has only three active dislocation systems, on the prism plane. The other two slip systems involve twinning. Short-ranged potentials typically have a low value for the basal stacking faults, and plasticity simulations using such potentials are dominated by basal slip, which would be incorrect for titanium. Correct stacking fault energies guarantee the correct dislocation behavior. This ensures that deformation will proceed by a combination of prism dislocation slip and twinning, as it does experimentally in Ti [48].

Special attention was paid to the determination of the phase transformation temperatures and the ability of the semi-empirical potentials to reproduce the experimental values. We found that the application of two absolutely independent techniques lead to agreement in the hcp-bcc transformation temperatures to within 2 K. The fitting to the phase transformation temperatures was found to be also rather straightforward within the EAM formalism. However, fitting the phase transformation data along with the T=0 target properties was found to be very challenging.

There seems to be a dichotomy between high-temperature thermodynamic properties, and low temperature static configurations. It is interesting that we are able to fit either set of properties accurately, but not both: this is also true of previous empirical models. We believe that our search for a suitable parameterization has been sufficiently comprehensive that this difference has a physical origin in properties beyond pair-functional potentials. An interatomic potential is, essentially, an attempt to integrate out the electronic degrees of freedom. The majority of temperature effects in metals arise from phonon entropy, and the interatomic potential responsible for these should not be temperature dependent. Temperature dependence of the electronic energy, due to the entropy of partially occupied electronic levels, is typically small; however it may play a crucial role in phase stability when two phases have very similar free energy. This has long been known to be the case in Ti and Zr, where the bcc structure has a very large electronic entropy. The entropy of the electrons is not captured explicitly by atomistic level simulations, so it should be included in the interatomic forces, i.e. the forces are derivatives of the electronic free energy, which in turn depends on the temperature. This effect can be incorporated by the Sommerfeld-type potentials [49].

Depending on the application, either high or low-T behaviour may be more important for a user. Consequently, we present two main alternatives: Potential Ti1 for high temperature transformations, and Ti2 for plasticity and point defect behavior in the hcp phase. Given that potential Ti1 works best at high temperatures and potential Ti2 at low temperatures, it is possible to produce a temperature-dependent Sommerfeld potential [49] as a linear combination of the two, which combines good point defect behavior at low temperatures with correct thermodynamic properties. Note that since we use the same function $\psi(r)$ in both potentials the linear combination of energies is identical to the linear combination of forces.

The Sommerfeld potential is in the EAM form, and has the same advantage of high speed enabling very large simulations. It is defined by introducing an interpolating function $g(T)$ which preserves the low temperature behavior up to room temperature and also retains the correct high temperature behavior. A suitable choice is a tanh function centered on an intermediate temperature:

$$g(T)=\tanh[(T-T_0)/T_w] \qquad (8)$$

with $T_0$=600 K, $T_w$=100 K. Then

$$\varphi_{Som}(T,r) = \{[1+g(T)]\,\varphi_{Ti1}(r) + [1-g(T)]\,\varphi_{Ti2}(r)\}/2 \qquad (9)$$
$$F_{Som}(T,\rho) = \{[1+g(T)]\,F_{Ti1}(\rho) + [1-g(T)]\,F_{Ti2}(\rho)\}/2 \qquad (10)$$

which means that we still have central forces acting between atoms i and j, which can be uniquely defined in pairwise terms as a linear combination of the forces of the two potentials. By construction, the T=0 properties of the Sommerfeld potential are essentially identical to those of Ti2. The hcp-bcc transformation temperature, volume and latent heat, and the melting point properties are those given by Ti1

The temperature $T$ which appears in this expression is the temperature of the electrons, not the kinetic energy of individual atoms. This means that the Sommerfeld potential can be used practically in two different ways. The first is to use an NPT or NVT ensemble with a thermostat, such as Nose-Hoover. This type of thermostat posits that the atomistic system is connected to a large thermal bath of constant temperature. The natural extension of this is to assume that the electrons are in equilibrium with the same bath. So the MD simulation is run with a fixed potential, generated at the temperature of the thermostat.

The Sommerfeld potential is trickier to implement if using the NPE or NVE ensembles. In these cases the electronic temperature changes, and the best assumption is to set it to the atomic temperature of the system. Consequently, at a given time-step, the potential is the same for all atoms; however it may change between time-steps. If the analytic form of the potential is used directly in the MD simulation program, this is not a problem, however if the code uses look-up tables these will need to be regenerated at each time-steps, which may prove prohibitive.

## 7. Conclusions

In the present study, three new EAM potentials for Ti were developed. Special attention was paid to determination of the phase transformation temperatures and the ability of the semi-empirical potentials to reproduce the experimental values. We found that the application of two absolutely independent techniques lead to agreement for the hcp-bcc transformation temperatures to within 2 K. Potential Ti1 well reproduces the hcp-bcc transformation and melting temperatures and is suitable for the simulation of bcc Ti. Potentials Ti2 and Ti3 dramatically underestimate the Ti melting temperature but correctly describe the hcp-bcc transformation temperature and can be used to simulate hcp Ti. The fact that a single EAM potential cannot describe both low and high temperature phases may be attributed to the different electronic structures of bcc and hcp titanium: notably bcc has a much higher electronic entropy. A Sommerfeld potential obtained from the combination of potentials Ti1 and Ti2 may be used to simulate the Ti properties at any temperature.

**Acknowledgements**
MIM's work (potential development and MD simulation) was supported by the U.S. Department of Energy (DOE), Office of Science, Basic Energy Sciences, Materials Science and Engineering Division. The research was performed at Ames Laboratory, which is operated for the U.S. DOE by Iowa State University under contract # DE-AC02-07CH11358. GJA acknowledges a Royal Society Wolfson Fellowship and Computer time from EPSRC grant EP/K014560 used for *ab* *initio* calculations. TLU's work (LSMC simulation) was supported by EPSRC grants eCSE04-4, EP/M011291/1, and a Doctoral Prize Fellowship.

Table I. Properties of crystal phases at T=0[¥].

| Property | Target value | Ti1 | Ti2 | Ti3 |
|---|---|---|---|---|
| $a$ (hcp) (Å) | 2.951 | **2.947** | **2.949** | **2.951** |
| $c/a$ (hcp) | 1.588 | **1.597** | **1.593** | **1.589** |
| $E_{coh}$ (eV/atom) | 4.85 | **5.346** | **5.247** | **5.402** |
| $C_{11}$ (hcp, GPa) | 176.1 [50] | **161** | **160** | **165** |
| $C_{12}$ (hcp, GPa) | 86.9 [50] | **80** | **70** | **88** |
| $C_{44}$ (hcp, GPa) | 50.8 [50] | **53** | **54** | **58** |
| $C_{13}$ (hcp, GPa) | 68.3 [50] | **86** | **70** | **83** |
| $C_{33}$ (hcp, GPa) | 172.5 [50] | **169** | **165** | **166** |
| $C_{66}$ (hcp, GPa) | 44.6 [50] | **40** | **45** | **39** |
| $\Delta E_{hcp \rightarrow \omega}$ (eV/atom) | 0.005 [30] | 0.005 | 0.007 | 0.016 |
| $a$ (fcc) (Å) | 4.115 [30] | 4.182 | 4.213 | 4.131 |
| $\Delta E_{hcp \rightarrow fcc}$ (eV/atom) | 0.059 [30] | 0.059 | 0.052 | 0.053 |
| $a$ (bcc) (Å) | 3.26 [30] | 3.251 | 3.256 | 3.242 |
| $\Delta E_{hcp \rightarrow bcc}$ (eV/atom) | 0.099 [30] | 0.029 | 0.074 | 0.089 |

---

[¥] The properties used in the fitting procedure are printed in bold.

Table II. Formation energies (eV) of point defects in hcp Ti at T=0[¥].

| Property | Target value [32] | Ti1 | Ti2 | Ti3 |
|---|---|---|---|---|
| $E_f^v$ | 1.97 | **2.74** | **1.78** | **1.73** |
| $E_f^O$ | 2.13 | **2.30** | **2.17** | **2.23** |
| $E_f^{BO}$ | 2.25 | **2.30** | **2.28** | **2.29** |
| $E_f^{BS}$ | 2.45 | **2.36** | **2.37** | **2.35** |
| $E_f^S$ | 2.48 | 3.15 | 2.79 | 3.01 |

---

[¥] The properties used in the fitting procedure are printed in bold.

Table III. Formation energies (mJ/m$^2$) of planar defects in hcp Ti at T=0[¥].

| Defect | Target value [51] | Ti1 | Ti2 | Ti3 |
|---|---|---|---|---|
| $I_1$ basal stacking fault defect energy | 148 | **130** | 119 | 118 |
| $I_2$ basal stacking fault defect energy | 259 | **257** | 236 | 236 |
| E basal stacking fault defect energy | 353 | **383** | 351 | 352 |
| Prism stacking fault defect energy | 250 | **208** | 257 | 255 |
| [0001] free surface energy | 124 | **86** | 120 | 141 |

---

[¥] The properties used in the fitting procedure are printed in bold.

Table IV. Phase transformation data[¥]. For the hcp-bcc transformation temperatures, the top value is obtained from MD simulation and the bottom values are obtained from the LSMC.

| Property | Target value [52] | Ti1 | Ti2 | Ti3 |
|---|---|---|---|---|
| $T_{\alpha \to \beta}$ (K) | 1155 | **1150±5** <br> 1152±3 | 1148±2 <br> 1150±2 <br> 1151±2 | 1148±2 <br> 1149±1 |
| $\Delta H_{\alpha \to \beta}$ (eV/atom) | 0.0435 | **0.022** | **0.032** | **0.041** |
| $\Delta V_{\alpha \to \beta}/V_\alpha$ (%) | | -0.70 | -0.03 | 0.71 |
| $T_m$ (hcp, K) | | 1765±1 | 1277±1 | 1189±1 |
| $\Delta H_m$ (hcp, eV/atom) | | 0.143 | 0.124 | 0.122 |
| $T_m$ (bcc, K) | 1941 | **1918±1** | 1322±1 | 1210±1 |
| $\Delta H_m$ (bcc, eV/atom) | 0.157 | **0.130** | 0.097 | 0.083 |

Table V. Activation energies for self-diffusion (eV/atom) in the vacancy (v) and self-interstitial mechanisms (i). For hcp Ti the point defect formation energies at T=1100 K were used and for the bcc the point defect formation energies at T=1500 K, T=1200 K and T=1100 K were used for potentials Ti1, Ti2 and Ti3, respectively.

| Phase | Experiment | *Ab initio* | Ti1 | Ti2 | Ti3 |
|---|---|---|---|---|---|
| hcp | 2.00 [53] <br> 3.14 [46] | v: 2.39 [44] <br> v: 2.61 [45] | v: 3.61 <br> i: 2.39 | v: 2.70 <br> i: 2.00 | v: 2.51 <br> i: 1.98 |
| bcc | 1.58 [54] <br> 1.35-2.60 [47] | | v: 2.86 <br> i: 1.51 | v: 1.76 <br> i: 0.84 | v: 1.25 <br> i: 0.70 |

Table VI. Point defect migration energies (eV/atom) obtained from MD simulation.

| Phase | Point defect | Ti1 | Ti2 | Ti3 |
|---|---|---|---|---|
| hcp | vacancy | 0.72±0.02 | 0.49±0.01 | 0.41±0.01 |
| | interstitial | 0.07±0.01 | 0.08±0.01 | 0.11±0.01 |
| bcc | vacancy | 0.30±0.01 | 0.24±0.01 | 0.24±0.01 |
| | interstitial | 0.11±0.01 | 0.10±0.01 | 0.13±0.01 |

---

[¥] The properties used in the fitting procedure are printed in bold.

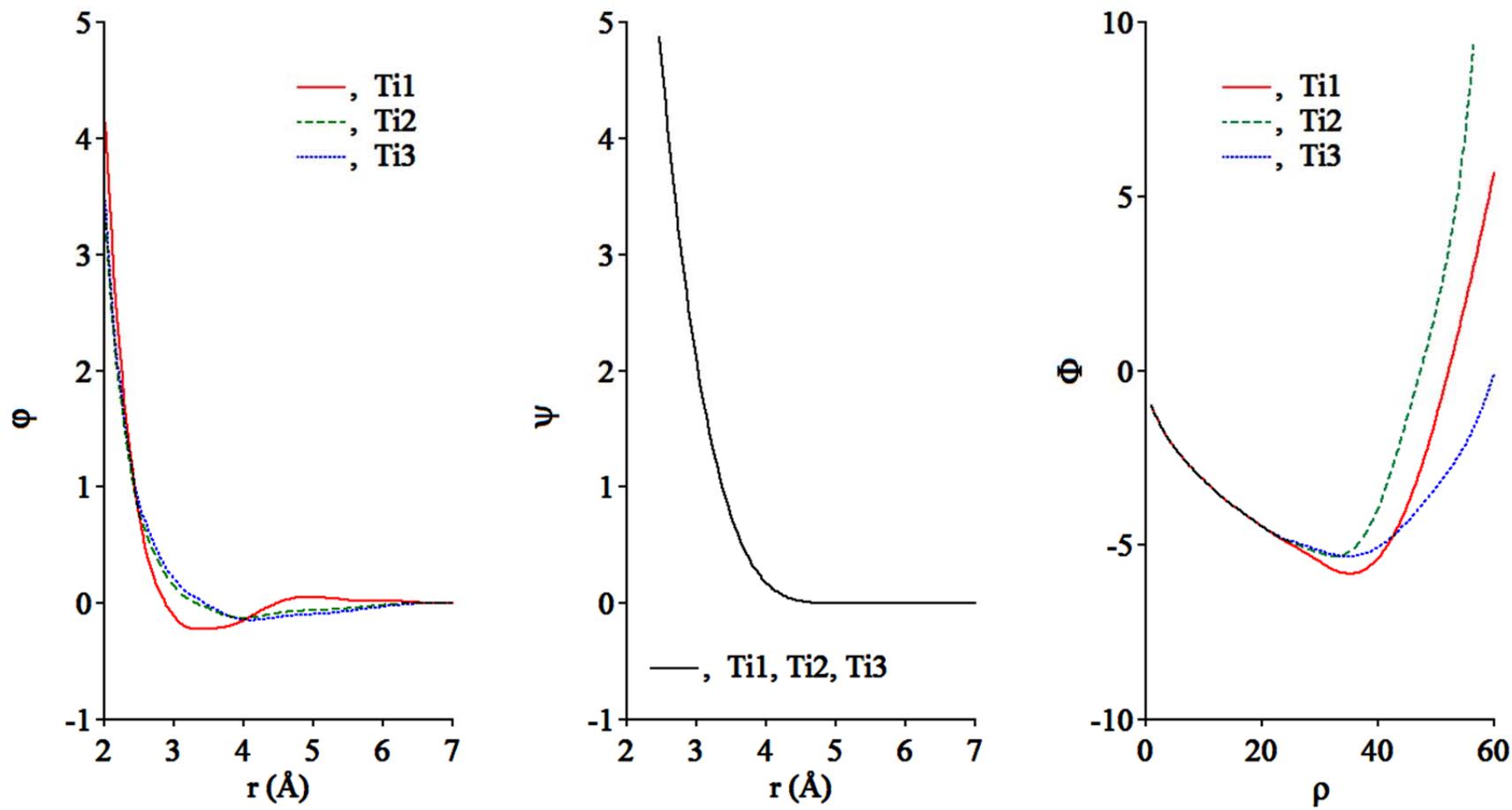

Figure 1. Potential functions.

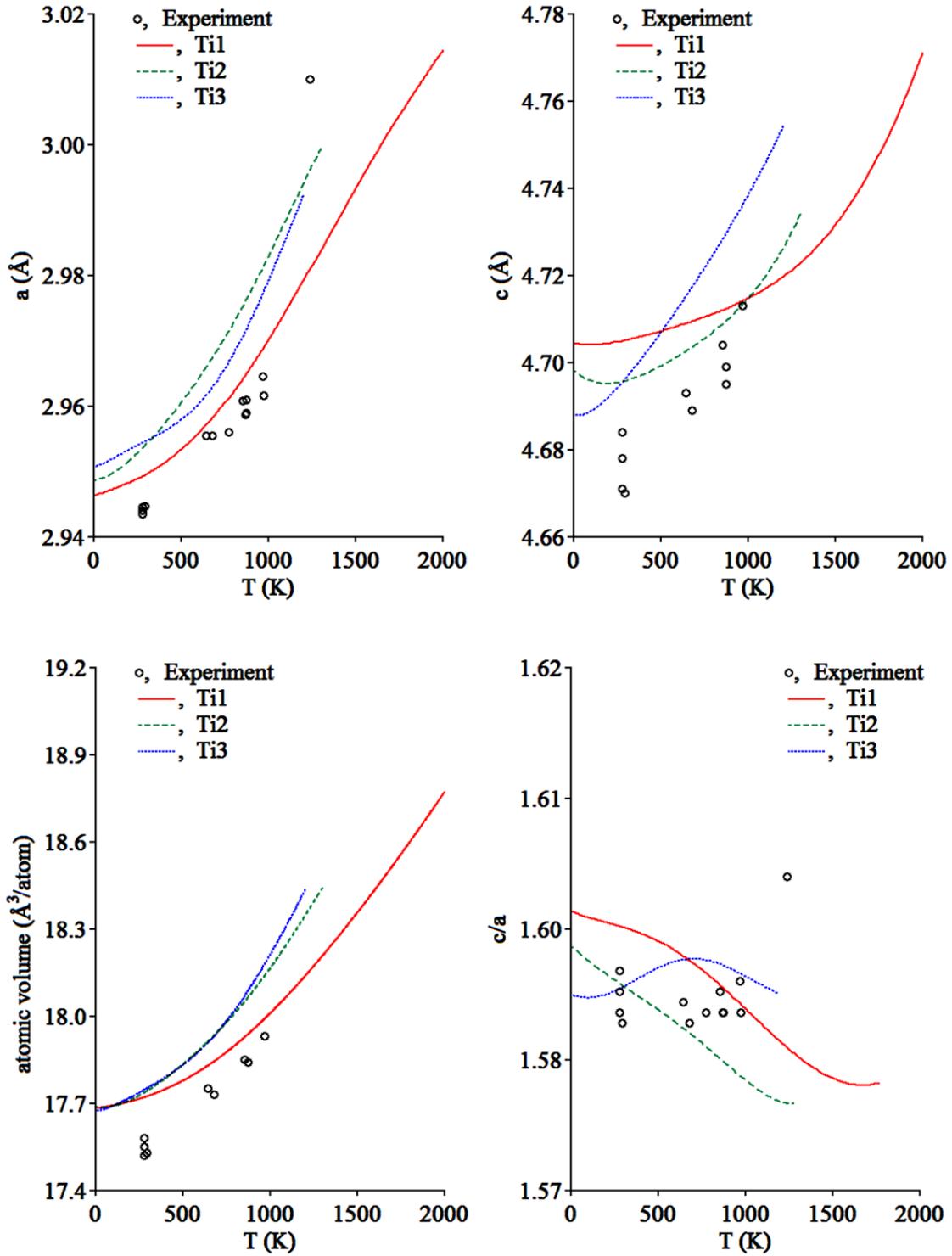

Figure 2. HCP lattice parameters as function of temperature. The experimental data are from [37].

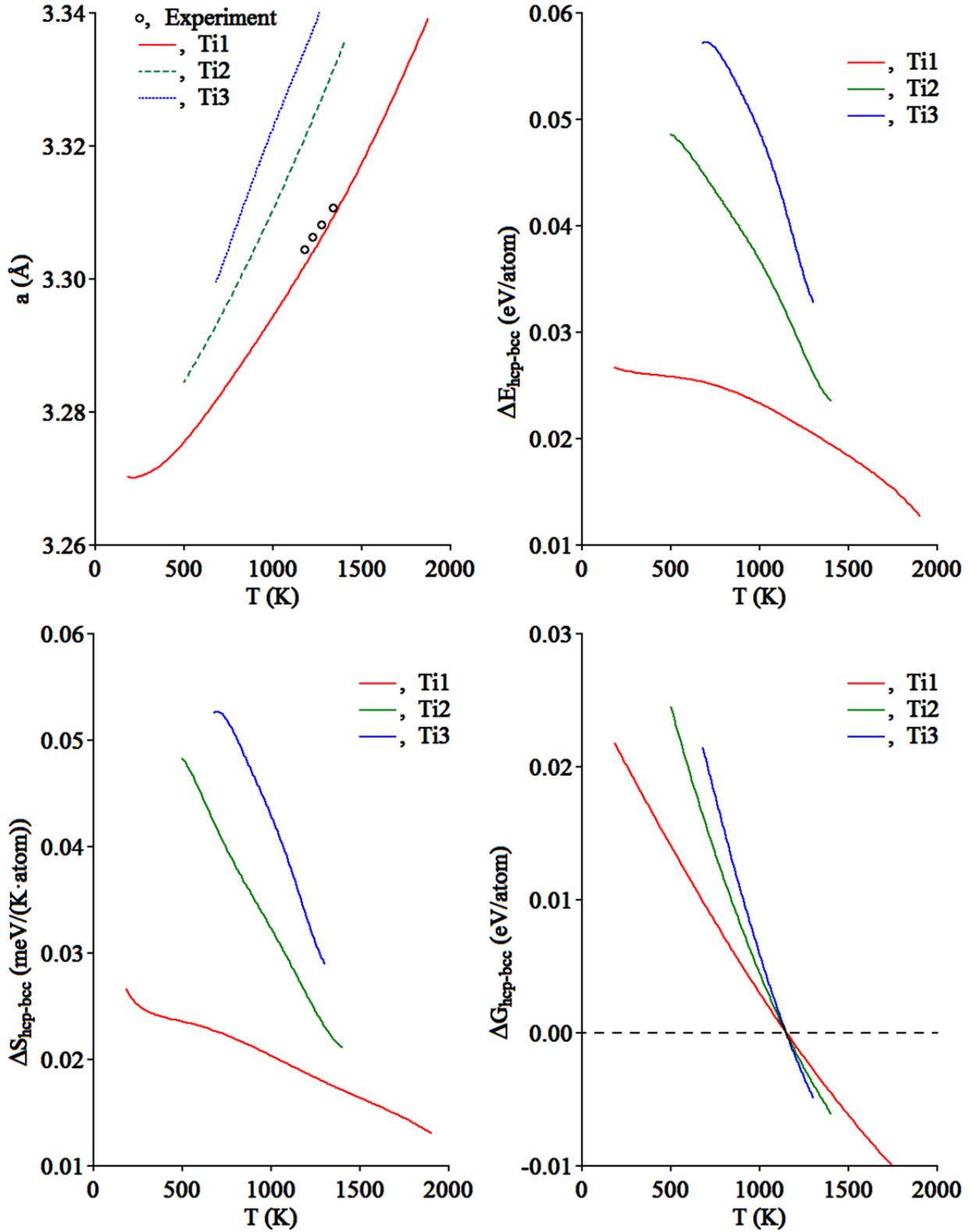

Figure 3. BCC lattice parameter, relative (to hcp) energy, entropy and free energy as functions of temperature. The experimental data are from [37].

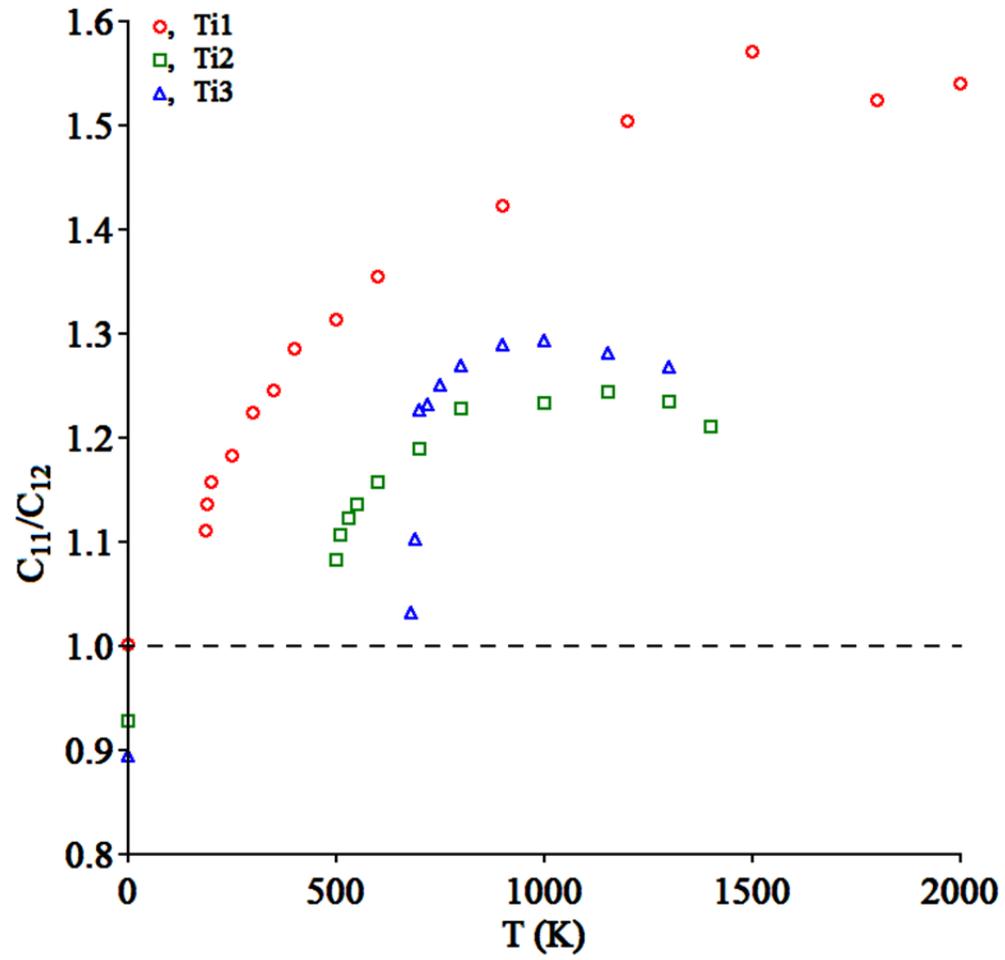

Figure 4. The $C_{11}/C_{12}$ ratio for bcc Ti as function of temperature.

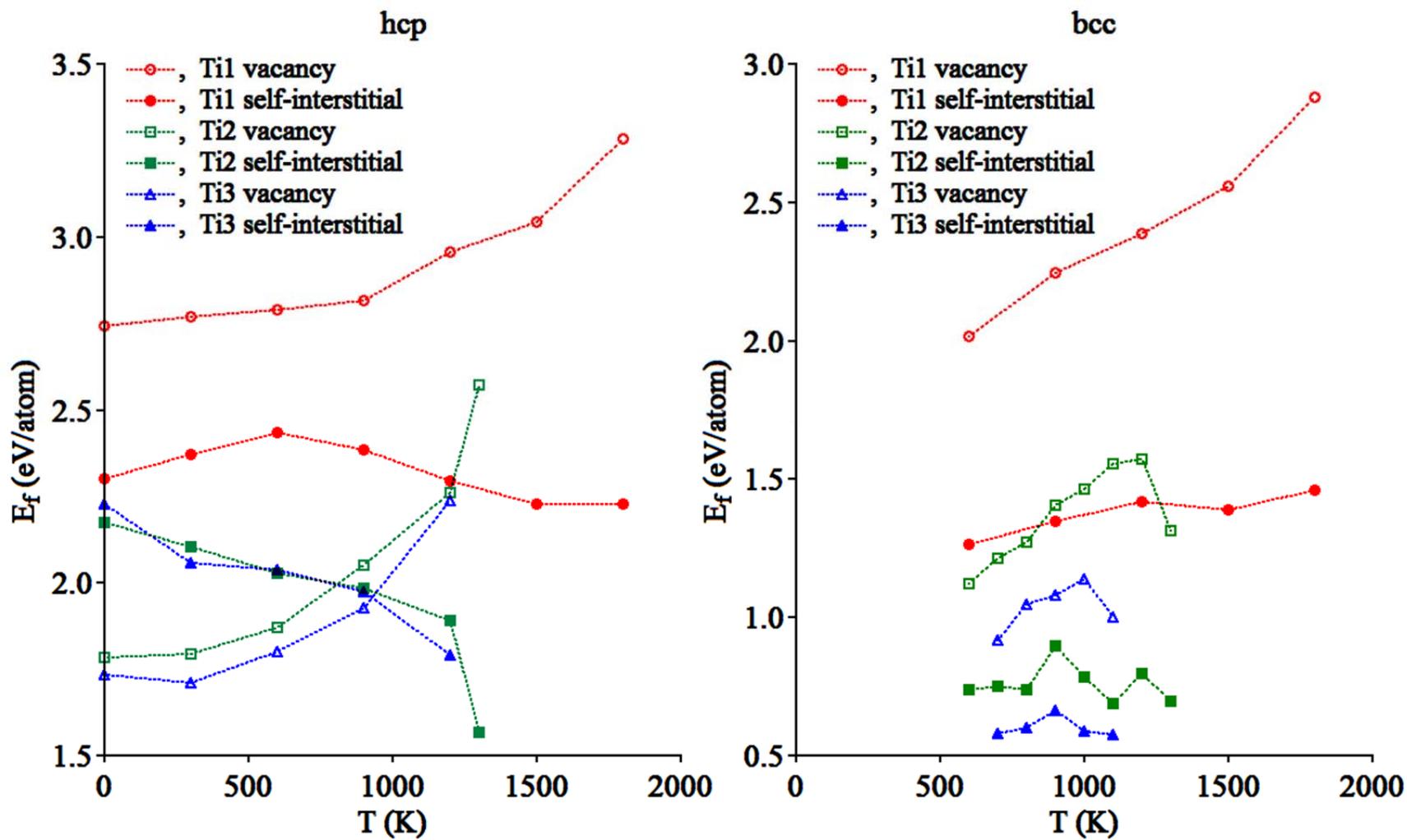

Figure 5. Point defect formation energies.

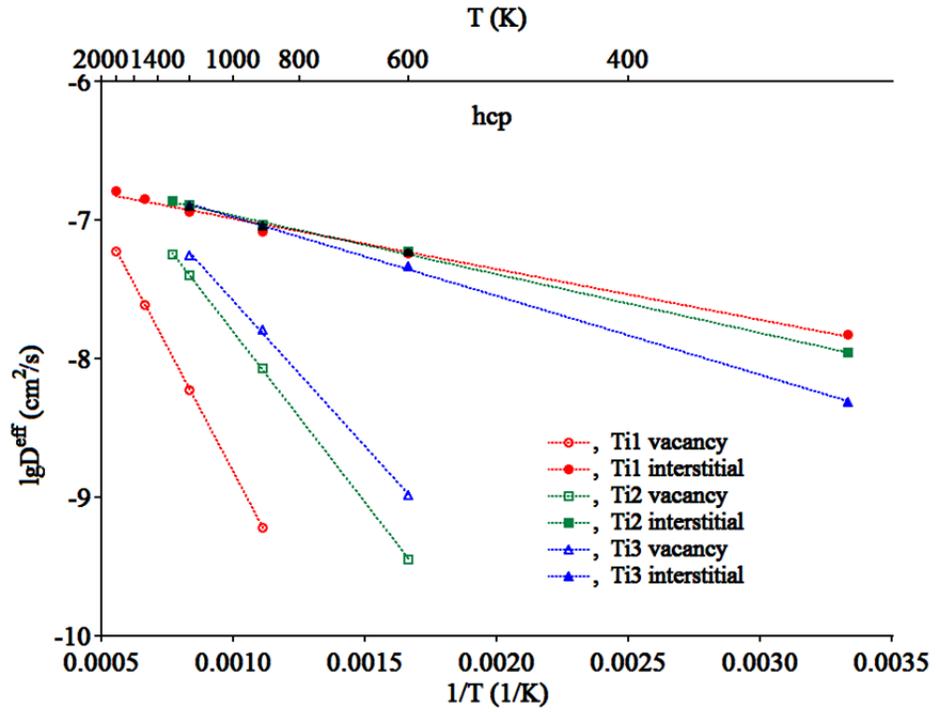

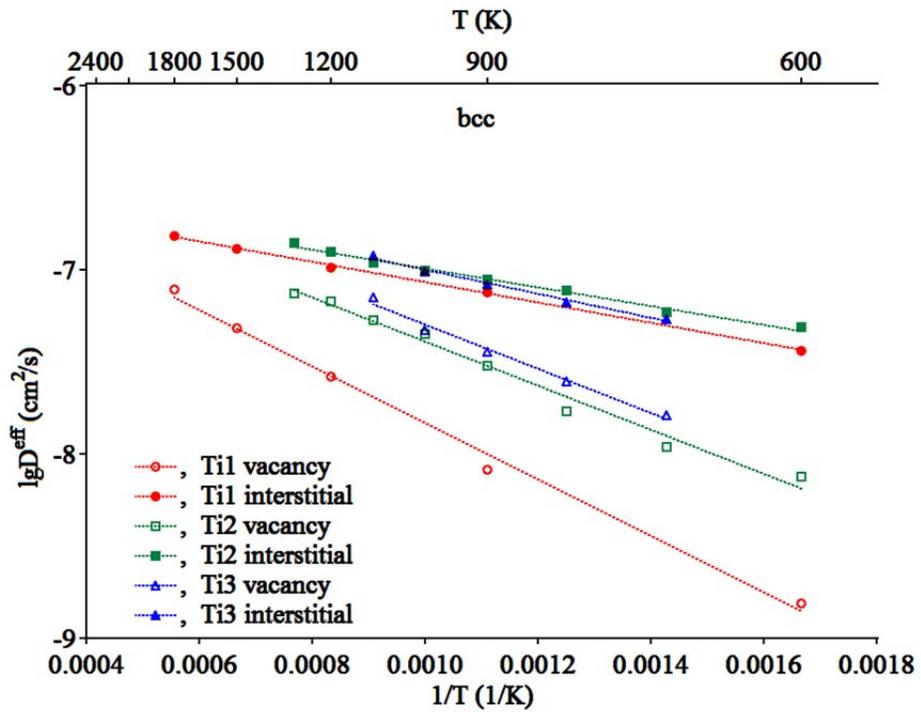

Figure 6. Effective diffusivities.

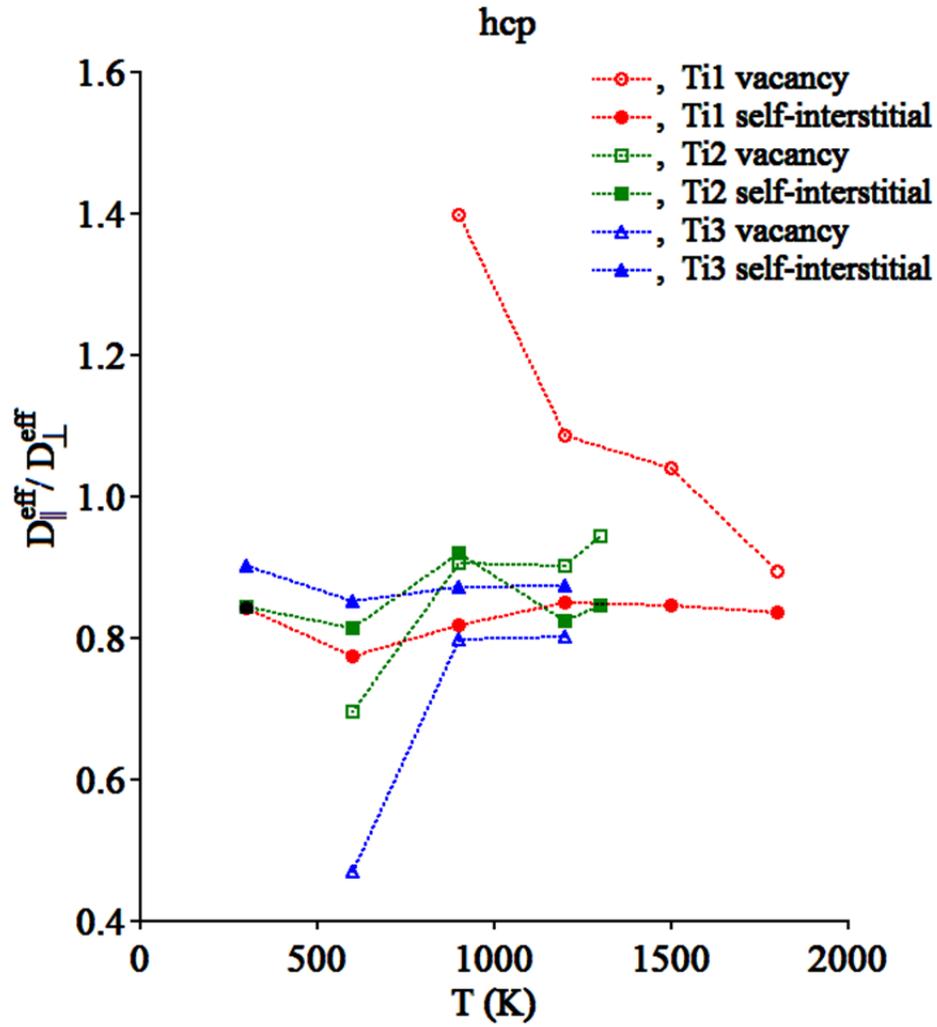

Figure 7. $D_{//}^{eff} / D_{\perp}^{eff}$ ratio for hcp Ti as function of temperature.